\documentclass[pra,print,showpacs,superscriptaddress,twocolumn]{revtex4}
\usepackage{amsmath}
\usepackage{amssymb}
\usepackage{graphicx}
\usepackage[english]{babel}
\usepackage{latexsym}
\usepackage{graphics}
\usepackage{subfigure}

\begin{document}

\title{BCS-BEC crossover in spin-orbit coupled two-dimensional Fermi gases}
\author{Gang Chen}
\affiliation{Department of Physics and Astronomy, Washington State University, Pullman,
Washington, 99164 USA}
\affiliation{Department of Physics, Shaoxing University, Shaoxing 312000, P. R. China}
\affiliation{State Key Laboratory of Quantum Optics and Quantum Optics Devices, College
of Physics and Electronic Engineering, Shanxi University, Taiyuan 030006, P.
R. China}
\author{Ming Gong}
\affiliation{Department of Physics and Astronomy, Washington State University, Pullman,
Washington, 99164 USA}
\author{Chuanwei Zhang}
\email{cwzhang@wsu.edu}
\affiliation{Department of Physics and Astronomy, Washington State University, Pullman,
Washington, 99164 USA}

\begin{abstract}
The recent experimental realization of spin-orbit coupling for ultra-cold
atoms has generated much interest in the physics of spin-orbit coupled
degenerate Fermi gases. Although recently the BCS-BEC crossover in
three-dimensional (3D) spin-orbit coupled Fermi gases has been intensively
studied, the corresponding two-dimensional (2D) crossover physics has
remained unexplored. In this paper, we investigate, both numerically and
analytically, the BCS-BEC crossover physics in 2D degenerate Fermi gases in
the presence of a Rashba type of spin-orbit coupling. We derive the mean
field gap and atom number equations suitable for the 2D spin-orbit coupled
Fermi gases and solve them numerically and self-consistently, from which the
dependence of the ground state properties (chemical potential, superfluid
pairing gap, ground state energy per atom) on the system parameters (e.g.,
binding energy, spin-orbit coupling strength) is obtained. Furthermore, we
derive analytic expressions for these ground state quantities, which agree
well with our numerical results within a broad parameter region. Such
analytic expressions also agree qualitatively with previous numerical
results for the 3D spin-orbit coupled Fermi gases, where analytic results
are lacked. We show that with an increasing SOC strength, the chemical
potential is shifted by a constant determined by the SOC strength. The
superfluid pairing gap is enhanced significantly in the BCS limit for strong
SOC, but only increases slightly in the BEC limit.
\end{abstract}

\maketitle

\section{Introduction}

Spin-orbit coupling (SOC), the interaction between the spin and orbital
degrees of freedom of a particle, has played important roles in condensed
matter as well as atomic and nuclear physics. For instance, it is known that
the coupling between electron spin and its linear momentum in solids leads
to many important condensed matter phenomena such as spin and anomalous Hall
effects \cite{NN,DX}, topological insulators and and superconductors \cite%
{Qi}, spintronics \cite{Zutic}, \textit{etc}. In atomic physics, the
coupling between the electron spin and its motion around an atomic nucleus
is responsible for many of the details of atomic structure \cite{Fine}.

Superfluidity and superconductivity are another important phenomena in
physics and have been widely studied in many physical systems, including
solids, Helium liquids, as well as ultra-cold atomic gases \cite{Leggett}.
Although particles in superfluids and superconductors usually possess spins
or pseudospins, the effects of SOC on superfluidity and superconductivity
have remained largely unexplored. In this context, the recent experimental
realization of SOC for ultra-cold atoms \cite{LYJ1} provides a completely
new platform for exploring many-body phenomena in spin-orbit coupled
superfluids, including both Bose-Einstein condensates (BECs) \cite%
{Zhai,allbose} and degenerate Fermi gases \cite{3DF,LD}. In the presence of
SOC, various new and exotic superfluid phenomena may exist since spins are
not conserved during their motion. In particular, in spin-orbit coupled
BECs, the ground state phase diagrams as well as the collective excitations
have been studied \cite{Zhai}. In spin-orbit coupled degenerate Fermi gases,
the crossover physics from the Bardeen-Cooper-Schrieffer (BCS) superfluids
of loosely bounded Cooper pairs to the BEC of tightly bounded molecules has
been investigated extensively, in both uniform and trapped three-dimensional
(3D) gases \cite{3DF,LD}. However, the study of spin-orbit coupled
two-dimensional (2D) Fermi gases is still lacked.

On the other hand, the 2D degenerate Fermi gas (without SOC) in itself is
one of the most active topics in ultra-cold atomic physics \cite{2DF}.
Experimentally, the 2D degenerate Fermi cold atomic gases have been realized
recently using a highly anisotropic pancake-shaped potential \cite%
{Martiyanov}, and the interaction energy in this system has been measured
using the radio-frequency spectroscopy \cite{BF}. In this system, tunable
interactions between atoms through Feshbach resonance \cite{FR} allow the
exploration of the crossover physics from the BCS superfluids to the BEC of
molecules in 2D \cite{MR}. In addition to the study of many-body physics
such as quantum fluctuations, non-Fermi liquid behavior, etc. \cite{Sachdev}%
, the 2D Fermionic cold atomic gases are particularly interesting because of
the existence of exotic topological excitations such as Majorana fermions
with non-Abelian exchange statistics \cite{Zhang}.

Motivated by these recent experimental breakthroughs in the realization of
SOC and 2D Fermi gases, in this paper, we investigate, both numerically and
analytically, the BCS-BEC crossover physics in 2D degenerate Fermi gases in
the presence of a Rashba type of SOC. Our main results are the following:

1) We derive the mean field gap and atom number equations suitable for the
2D spin-orbit coupled Fermi gases.

2) These two equations are solved numerically and self-consistently, from
which the dependence of the ground state properties (chemical potential,
superfluid pairing gap, ground state energy per atom) on the system
parameters (e.g., binding energy, SOC strength) is obtained.

3) We derive analytic expressions for these ground state physical
quantities, which agree well with our numerical results within a broad
parameter region. Such analytic expressions also agree qualitatively with
previous numerical results for the 3D spin-orbit coupled Fermi gases \cite%
{3DF}, where analytic results are lacked.

4) We find that with an increasing SOC strength $\alpha $, the chemical
potential is shifted by a constant $-m\alpha ^{2}$ determined by $\alpha $.
The superfluid pairing gap and the ground state energy per atom are affected
at the order of $\alpha ^{4}$ rather than $\alpha ^{2}$, which means that
weak SOC does not affect the pairing gap and the ground state energy per
atom. In the strong SOC regime, the pairing gap and the ground state energy
are enhanced significantly in the BCS limit, but only increase slightly in
the BEC limit. Although these analytic results are obtained for the 2D Fermi
gases, they also provide qualitative understanding for the numerical results
in 3D Fermi gases \cite{3DF}, where similar changes of the chemical
potential and superfluid pairing order with respect to the SOC strength are
observed but analytic results are lacked.

The paper is organized as follows. Section II describes the physical system:
the spin-orbit coupled 2D degenerate Fermi gases, and the corresponding
Hamiltonian. In section III, we derive the mean field gap and atom number
equations. These equations are self-consistently solved in section IV, both
numerically and analytically (through perturbative methods), to obtain the
ground state properties (chemical potential, superfluid order parameter, and
ground state energy per atom) of the spin-orbit coupled Fermi gases in the
BCS-BEC crossover. Section V consists of discussion and conclusion.

\section{The physical system and the Hamiltonian}

The physical system in consideration is a 2D degenerate Fermi gas.
Experimentally, the 2D degenerate Fermi gas has been realized using a 1D
deep optical lattice along the third dimension, where the tunneling between
different layers is suppressed completely \cite{Martiyanov,BF}. The 1D
optical lattice potential $V_{0}\sin ^{2}(2\pi z/\lambda _{w})$ can be
generated using two counter-propagating laser beams (parallel to the $z$
axis with a wavelength $\lambda _{w}$). In this case, the two-body binding
energy is given by $E_{b}\simeq 0.915\hbar \omega _{L}\exp (\sqrt{2\pi }%
l_{L}/a_{s})/\pi $, where $\omega _{L}=\sqrt{8\pi ^{2}V_{0}/(m\lambda
_{w}^{2})}$ is the effective trapping frequency along the $z$ axis, $l_{L}=%
\sqrt{\hbar /(m\omega _{L})}$, and $a_{s}$ is the 3D \textit{s}-wave
scattering length \cite{JT}. Therefore the two-body binding energy $E_{b}$
can be tuned by varying the \textit{s}-wave scattering length $a_{s}$ via
the Feshbash resonance for the study of the BCS-BEC crossover physics \cite%
{MR}. For a small attractive $a_{s}\rightarrow 0^{-}$, $E_{b}\rightarrow 0$,
corresponding to the BCS limit. While for a small repulsive $%
a_{s}\rightarrow 0^{+}$, $E_{b}\rightarrow \infty $, corresponding to the
BEC limit. When $E_{b}$ increases from 0 to $\infty $, the system evolves
continuously from a BCS superfluid to a BEC of molecules.

The SOC for cold atoms can be generated by the interaction between atoms and
laser beams, as shown in many previous literatures \cite{lasersetup}, and
demonstrated in a recent benchmark experiment \cite{LYJ1}. In this paper, we
consider only a Rashba type of SOC, and the effects of other types of SOC
(e.g. Dresselhaus or the combination of both) can be investigated similarly.
For simplicity we consider a uniform Fermi gas and neglect the weak harmonic
trap in the 2D plane, whose effects can be incorporated using the local
density approximation.

The Hamiltonian for this uniform 2D spin-orbit coupled degenerate Fermi
gases can be written as%
\begin{equation}
H=H_{\text{F}}+H_{\text{I}}+H_{\text{soc}},  \label{TH}
\end{equation}%
where
\begin{equation}
H_{\text{F}}=\sum_{\mathbf{k},\sigma }\zeta _{\mathbf{k}}C_{\mathbf{k}\sigma
}^{\dagger }C_{\mathbf{k}\sigma }  \label{HF}
\end{equation}%
is the single atom Hamiltonian, $C_{\mathbf{k}\sigma }^{\dagger }$ is the
creation operator for a Fermi atom with the momentum $\mathbf{k}=\left(
k_{x},k_{y}\right) $, $\sigma =\uparrow ,\downarrow $ are the pseudospins of
atoms. $\zeta _{\mathbf{k}}=\epsilon _{\mathbf{k}}-\mu $ with the kinetic
energy $\epsilon _{\mathbf{k}}=k^{2}/2m$, the chemical potential $\mu $, and
the atom mass $m$. Henceforth, we take the Planck constant $\hbar =1$. The
s-wave scattering interaction between atom
\begin{equation}
H_{\text{I}}=g\sum_{\mathbf{k}}C_{-\mathbf{k}\uparrow }^{\dagger }C_{\mathbf{%
k}\downarrow }^{\dagger }C_{\mathbf{k}\downarrow }C_{-\mathbf{k}\uparrow },
\label{HI}
\end{equation}%
where $g$ is the effective scattering interaction parameter. In a 2D Fermi
gas,
\begin{equation}
\frac{1}{g}=-\sum_{\mathbf{k}}\frac{1}{(2\epsilon _{\mathbf{k}}+E_{b})}.
\label{PG}
\end{equation}

The Hamiltonian for the Rashba type of SOC for atoms can be written as
\begin{equation}
H_{\text{soc}}=\alpha \sum_{\mathbf{k}}[(k_{y}+ik_{x})C_{\mathbf{k}\uparrow
}^{\dagger }C_{\mathbf{k}\downarrow }+(k_{y}-ik_{x})C_{\mathbf{k}\downarrow
}^{\dagger }C_{\mathbf{k}\uparrow }],  \label{HSOC}
\end{equation}%
where $\alpha $ is the SOC strength. In solid state materials, $\alpha $ is
generally much smaller than $K_{F}/(2m)$ with $K_{F}$ as the Fermi vector.
However, in ultra-cold neutral atoms, $\alpha $ can reach the order of $%
K_{F}/(2m)$ \cite{lasersetup}. Such strong SOC, together with tunable
interactions through the Feshbach resonance, may yield some exotic many-body
phenomena that have not been explored in solid state systems. Recently, a
generalized SOC with the Rashba and the Dresselhaus terms in the Hamiltonian
(\ref{TH}) has been considered \cite{LD}.

\section{Mean field gap and atom number equations}

As the first step for the eventual understanding of the 2D
spin-orbit coupled Fermi gas, we consider the zero temperature
superfluid physics under the mean field approximation \cite{MR}.
Generally, the mean field approximation can give qualitatively but
not quantitatively correct results \cite{Bertaina}. In the mean
field approximation, the superfluid order parameter is taken as
\begin{equation}
\Delta =g\sum_{\mathbf{k}}\left\langle C_{\mathbf{k}\downarrow }C_{-\mathbf{k%
}\uparrow }\right\rangle .  \label{order}
\end{equation}%
With this pairing order parameter, we can rewrite the two-body interaction
Hamiltonian (\ref{HI}) as
\begin{equation}
H_{\text{I}}=-\Delta ^{2}/g+\Delta \sum_{\mathbf{k}}(C_{\mathbf{k}\downarrow
}C_{-\mathbf{k}\uparrow }+C_{-\mathbf{k}\uparrow }^{\dagger }C_{\mathbf{k}%
\downarrow }^{\dagger }).  \label{interactionH}
\end{equation}%
Therefore the total Hamiltonian can be rewritten as
\begin{equation}
H_{\text{B}}=\frac{1}{2}\sum_{\mathbf{k}}\Psi ^{\dagger }(\mathbf{k})M_{%
\mathbf{k}}\Psi (\mathbf{k})-\frac{\Delta ^{2}}{g}+\sum_{\mathbf{k}}\zeta _{%
\mathbf{k}}  \label{BDG}
\end{equation}%
under the Nambu spinor basis $\Psi (\mathbf{k})=(C_{\mathbf{k}\uparrow },C_{%
\mathbf{k}\downarrow },C_{-\mathbf{k\downarrow }}^{\dagger },-C_{-\mathbf{k}%
\uparrow }^{\dagger })^{T}$, where Bogoliubov-de-Gennes operator
\begin{equation}
M_{\mathbf{k}}=\left(
\begin{array}{cccc}
\zeta _{\mathbf{k}} & \alpha k_{+} & \Delta  & 0 \\
\alpha k_{-} & \zeta _{\mathbf{k}} & 0 & \Delta  \\
\Delta  & 0 & -\zeta _{\mathbf{k}} & -\alpha k_{+} \\
0 & \Delta  & -\alpha k_{-} & -\zeta _{\mathbf{k}}%
\end{array}%
\right)   \label{MAT}
\end{equation}%
preserves the particle-hole symmetry, $k_{\pm }=k_{y}\pm ik_{x}$.

The quasiparticle excitation spectrum
\begin{equation}
E_{\mathbf{k},\pm }^{\lambda }=\lambda \sqrt{(\epsilon _{\mathbf{k}}-\mu \pm
\alpha k)^{2}+\Delta ^{2}}  \label{ESE}
\end{equation}%
is the eigenvalue of the matrix $M_{\mathbf{k}}$, $\lambda =\pm $ correspond
to the particle and hole branches of the spectrum. For each branch, there
are two different excitations due to the existence of the SOC. From Eq. (\ref%
{BDG}), we see the total ground-state energy is
\begin{equation}
E_{G}=-\frac{\Delta ^{2}}{g}+\sum\nolimits_{\mathbf{k}}[\zeta _{\mathbf{k}}-%
\frac{1}{2}(E_{\mathbf{k},+}^{+}+E_{\mathbf{k},-}^{+})].  \label{GTE}
\end{equation}%
Without the SOC, the term $(E_{\mathbf{k},+}^{+}+E_{\mathbf{k},-}^{+})/2$
reduces to the well-known form
\begin{equation}
E_{\mathbf{k}}=\sqrt{(\epsilon _{\mathbf{k}}-\mu )^{2}+\Delta ^{2}}.
\label{quasi2}
\end{equation}%
in the BCS theory.

The ground-state properties of the 2D spin-orbit coupled Fermi gases can be
obtained from the atom number equation
\begin{equation}
n=-\frac{\partial E_{G}}{\partial \mu }=\sum\nolimits_{\mathbf{k}}[1+\frac{1%
}{2}(\frac{\partial E_{\mathbf{k}+}^{+}}{\partial \mu }+\frac{\partial E_{%
\mathbf{k}-}^{+}}{\partial \mu })],  \label{number}
\end{equation}%
and the superfluid gap equation
\begin{equation}
\frac{\partial E_{G}}{\partial \Delta }=\sum_{\mathbf{k}}[\frac{\Delta }{%
2\epsilon _{\mathbf{k}}+E_{b}}-\frac{1}{4}(\frac{\partial E_{\mathbf{k}+}^{+}%
}{\partial \Delta }+\frac{\partial E_{\mathbf{k}-}^{+}}{\partial \Delta }%
)]=0.  \label{gap}
\end{equation}

\section{Ground state properties}

\subsection{Numerical results}

We numerically solve the above atom number equation (\ref{number}) and the
gap equation (\ref{gap}) self-consistently to obtain various ground state
quantities. In Fig. 1, we plot the dependence of the ground state
quantities: the chemical potential $\mu $, the superfluid order parameter $%
\Delta $, and the ground state energy per atom $E=E_{G}/n$, on the physical
parameters: the SOC strength $\alpha $ and the binding energy $E_{b}$. We
see that the chemical potential $\mu $ decreases with the increasing
spin-orbit coupling strength $\alpha $. With increasing binding energy, the
chemical potential also decreases, signaling the crossover physics from the
BCS superfluids to the BEC molecules. One interesting feature shown in Fig.
1(a2) is that the shift of the chemical potential induced by the SOC depends
only on the SOC strength.

\begin{figure}[bp]
\includegraphics[width = 1\linewidth]{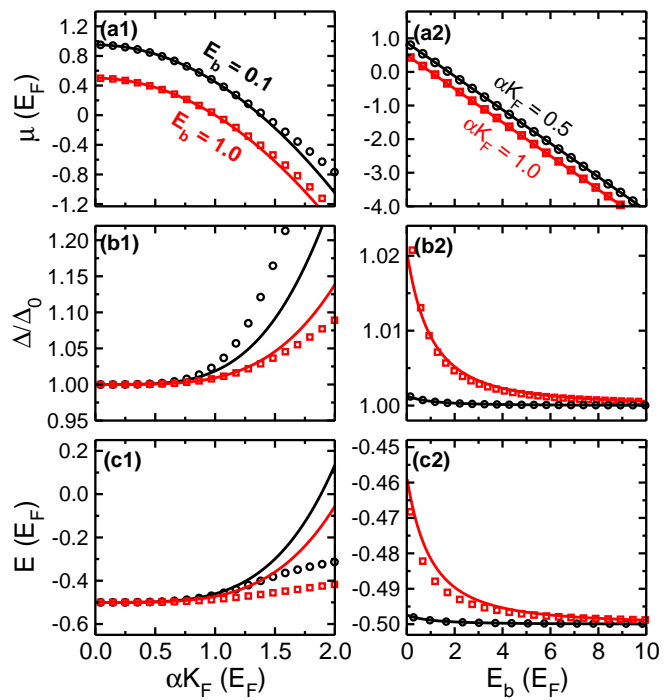}
\caption{(Color online) (a1-c1) Plot of the chemical potential $\protect\mu $%
, the dimensionless superfluid pairing gap $\Delta /\Delta _{0}$, and the
ground-state energy per atom $E$ with respect to the SOC strength $\protect%
\alpha $ for the two-body binding energy $E_{b}=0.1E_{F}$ (Black line) and $%
1.0$ $E_{F}$ (Red line). (a2-c2) Plot of $\protect\mu $, $\Delta /\Delta
_{0} $, and $E$ with respect to $E_{b}$ for $\protect\alpha K_{F}=0.5E_{F}$
(Black line) and $1.0$ $E_{F}$ (Red line). In all figures, the solid lines
represent the approximate analytical results obtained in the paper, while
the open symbols correspond to the exact numerical results.}
\label{fig1}
\end{figure}

In Fig. 1(b1), we see the superfluid order parameter $\Delta $ increases
with increasing $\alpha $. Here $\Delta _{0}$ is the superfluid order
parameter without SOC. For a small $\alpha $, the change of $\Delta /\Delta
_{0}$ is very small. The growth of $\Delta $ becomes significant only when $%
\alpha K_{F}$ is large than the Fermi energy $E_{F}$. Furthermore, the
effects of the SOC are clearly seen for a small $E_{b}$ (the BCS side), but
not important for a large $E_{b}$ (the BEC side).

The ground state energy per atom $E$ has a similar behavior as $\Delta
/\Delta _{0}$. It increases with $\alpha $, but the growth is only important
for a large $\alpha $. Without SOC $E=-\frac{1}{2}E_{F}$, and the SOC
induces a small correction. The change of $E$ is significant in the BCS side
(small $E_{b}$), but negligible in the BEC side (large $E_{b}$). We note
that similar changes of the chemical potential and the superfluid order
parameter have been observed numerically in a 3D spin-orbit coupled Fermi
gases. All these numerical observed phenomena will be explained in the next
subsection where the analytic expressions for these ground state quantities
are derived.

\subsection{Analytic results}

Although the above numerical results give the dependence of the ground state
quantities on the SOC strength for certain parameters, analytic results are
desired for a better understanding of the underlying physics. In the
following, we present a perturbative method (with $\alpha $ as the small
parameter) to analytically extract the fundamental ground-state properties
of the spin-orbit coupled Fermi gases. For this purpose, we rearrange Eq. (%
\ref{GTE}) into two parts%
\begin{equation}
E_{G}=E_{0}+E_{\text{soc}}  \label{GTE-1}
\end{equation}%
with
\begin{equation}
E_{0}=-\frac{\Delta ^{2}}{g}+\sum\nolimits_{\mathbf{k}}(\zeta _{\mathbf{k}%
}-E_{\mathbf{k}})  \label{GW}
\end{equation}%
as the ground-state energy without SOC. For a 2D Fermi gas, $E_{0}$ can be
obtained exactly, yielding \cite{MR}
\begin{equation}
E_{0}=\frac{m}{4\pi }[\Delta ^{2}\ln (\frac{\sqrt{\mu ^{2}+\Delta ^{2}}-\mu
}{E_{b}})-\mu (\sqrt{\mu ^{2}+\Delta ^{2}}+\mu )-\frac{\Delta ^{2}}{2}].
\label{GTWS}
\end{equation}%
The second term in Eq. (\ref{GTE-1}) is given by
\begin{equation}
E_{\text{soc}}=\sum\nolimits_{\mathbf{k}}[E_{\mathbf{k}}-\frac{1}{2}(E_{%
\mathbf{k},+}^{+}+E_{\mathbf{k},-}^{+})],  \label{GWO}
\end{equation}%
which describes the contribution from the SOC.

Because it is difficult to derive a simple analytic expression for $E_{\text{%
soc}}$ directly, we first perform a Taylor expansion with respect to the SOC
strength, and then do the summation over $\mathbf{k}$. Formally, $E_{\text{%
soc}}$ can be written as%
\begin{equation}
E_{\text{soc}}=\frac{m}{4\pi }C_{i}(\mu ,\Delta )\eta ^{i}  \label{ESO}
\end{equation}%
with $\eta =m\alpha ^{2}/2$. The coefficients $C_{i}(\mu ,\Delta )$ can be
obtained, in principle, for any order. Here, we only give the first six
orders
\begin{eqnarray}
C_{1} &=&-4(\sqrt{\mu ^{2}+\Delta ^{2}}+\mu ),  \label{C1} \\
C_{2} &=&-\frac{8}{3}(1+\frac{\mu }{\sqrt{\mu ^{2}+\Delta ^{2}}}), \\
C_{3} &=&\frac{-16\Delta ^{2}}{15(\mu ^{2}+\Delta ^{2})^{3/2}}, \\
C_{4} &=&\frac{32\mu \Delta ^{2}}{35(\mu ^{2}+\Delta ^{2})^{5/2}}, \\
C_{5} &=&\frac{64\Delta ^{2}(\Delta ^{2}-4\mu ^{2})}{315(\mu ^{2}+\Delta
^{2})^{7/2}}, \\
C_{6} &=&\frac{-128\mu \Delta ^{2}(3\Delta ^{2}-4\mu ^{2})}{693(\mu
^{2}+\Delta ^{2})^{9/2}}.
\end{eqnarray}

Although the expression for $E_{\text{soc}}$ is very complicate, the
expressions for the chemical potential $\mu $ and the superfluid pairing
order $\Delta $ are very simple, as we will show later in the paper. With
the ground-state energy $E_{G}$, the superfluid pair order and the chemical
potential can be derived by self-consistently solving the corresponding gap
and number equations
\begin{eqnarray}
\ln (\frac{\sqrt{\mu ^{2}+\Delta ^{2}}-\mu }{E_{b}}) &=&G_{\Delta }(\eta
,\Delta ,\mu ),  \label{Gap1} \\
\sqrt{\mu ^{2}+\Delta ^{2}}+\mu &=&2E_{F}-G_{\mu }(\eta ,\Delta ,\mu )
\label{CP1}
\end{eqnarray}%
analytically, where $G_{\Delta }(\eta ,\Delta ,\mu )=\sum_{i}\frac{\partial
C_{i}(\mu ,\Delta )}{\partial \Delta }\eta ^{i}$, and $G_{\mu }(\eta ,\Delta
,\mu )=\sum_{i}\frac{\partial C_{i}(\mu ,\Delta )}{\partial \mu }\eta ^{i}$.
$E_{F}$ is the Fermi energy for a 2D non-interacting Fermi gas without SOC
and with the density $n=mE_{F}/\pi $. Without SOC ($\eta =0$), $G_{\Delta
}(\eta ,\Delta ,\mu )=G_{\mu }(\eta ,\Delta ,\mu )=0$, and Eqs. (\ref{Gap1})
and (\ref{CP1}) become
\begin{eqnarray*}
\sqrt{\mu ^{2}+\Delta ^{2}}-\mu &=&E_{b}, \\
\sqrt{\mu ^{2}+\Delta ^{2}}+\mu &=&2E_{F}.
\end{eqnarray*}%
In this case, the superfluid pairing order and the chemical potential are
given exactly by $\Delta _{0}=\sqrt{2E_{b}E_{F}}$ and $\mu
_{0}=E_{F}-E_{b}/2 $ \cite{MR}, as expected.

In the presence of SOC ($\eta \neq 0$), the nonlinear equations (\ref{Gap1})
and (\ref{CP1}) cannot be solved exactly. However, approximate solutions can
be derived for the physical parameters within current experimentally
achievable region. Since the ground-state energy depends on $\Delta ^{2}$,
the solutions of Eqs. (\ref{Gap1}) and (\ref{CP1}) can be assumed to be $\mu
=\sum_{i=0}\mu _{i}\eta ^{i}$ and $\Delta ^{2}=\sum_{i=0}\Lambda _{i}\eta
^{i}$. Substituting these expressions into the nonlinear equations (\ref%
{Gap1}) and (\ref{CP1}) and then comparing the coefficients for the same
order of $\eta ^{i}$, we obtain $\Lambda _{i}$ and $\mu _{i}$ respectively.

\subsubsection{Chemical potential}

After a straightforward but tedious calculation, the coefficients for the
chemical potential are given by $\mu _{1}=-2$, $\mu
_{2}=4E_{b}/[3(E_{b}+2E_{F})^{2}]$ and $\mu
_{3}=-64E_{b}(E_{b}-4E_{F})/[15(E_{b}+2E_{F})^{4}]$. Note that the second
order $\mu _{2}$ is already very small for all different values of $E_{b}$
when $\eta <E_{F}$, therefore the high-order terms do not affect the
chemical potential significantly. The chemical potential can then be written
as
\begin{equation}
\mu \simeq E_{F}-\frac{E_{b}}{2}-2\eta .  \label{MSTC}
\end{equation}%
From Fig. 1a, we see Eq. (\ref{MSTC}) agrees well with the exact values of
the chemical potential obtained from numerically solving Eqs. (\ref{number})
and (\ref{gap}) self-consistently.

\begin{figure}[bp]
\includegraphics[width = 1\linewidth]{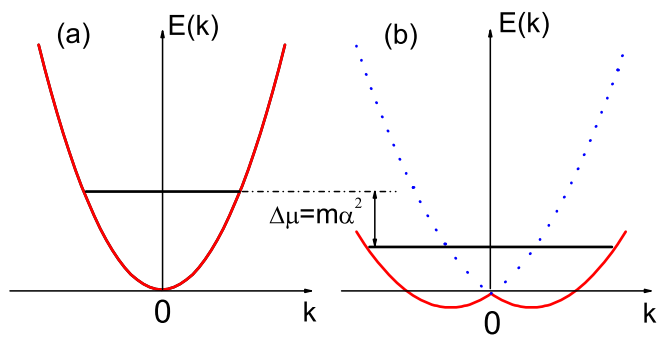}
\caption{(Color online) (a) The chemical potential without SOC in the BCS
limit. (b) The chemical potential with SOC in the BCS limit. For weak SOC,
their difference $\Delta \protect\mu =m\protect\alpha ^{2}$.}
\label{fig2}
\end{figure}
Equation (\ref{MSTC}) shows clearly that with the increasing SOC strength $%
\alpha $, the chemical potential $\mu $ is decreased by $2\eta =m\alpha ^{2}$%
, as shown in Fig. 2. If we define an effective chemical potential $%
\overline{\mu }=\mu +2\eta =\mu +m\alpha ^{2}$, Eq. (\ref{MSTC}) can be
rewritten as $\overline{\mu }=E_{F}-E_{b}/2$. Similar to the previous
discussion without SOC, we find that in the asymptotic BCS limit with a weak
bound state ($E_{b}\ll E_{F}$), the chemical potential $\overline{\mu }%
\simeq E_{F}$. However, in the deep BEC regime with a strong bound state ($%
E_{b}\gg E_{F}$), the chemical potential $\overline{\mu }=-E_{b}/2$. The
BCS-BEC crossover may occur around the crossover point $\overline{\mu }=0$
\cite{MR}. Furthermore, in the presence of SOC, Eq. (\ref{MSTC}) can also be
written as $\mu =\mu _{f}+\mu _{\Delta }$, where $\mu _{f}=E_{F}-2\eta $ is
the chemical potential for the free Fermi gas and $\mu _{\Delta
}=-E_{b}/2+\sum_{i=1}\mu _{\Delta i}(E_{b})\eta ^{i}$ reflects the revision
of the chemical potential induced by the two-body binding energy $E_{b}$.
Here $\mu _{\Delta 1}=0$ implies that the binding energy $E_{b}$ has no
effect on the chemical potential $\mu $ at the order of $\eta =m\alpha
^{2}/2 $.

It is important to point out that although the chemical potential $\mu $
only has a simple shift from that without SOC, the underlying physics is
quite different. Without SOC, the pairing wave function is simply singlet.
However, in the presence of SOC, each energy band contains both spin up and
down components. As a result, the pairing wave function has a more
complicated structure with both singlet and triplet components \cite{LPG}.
The triplet pairing correlations in \textit{s}-wave superfluids may be used
to detect the anisotropic Fulde-Ferrell-Larkin-Ovchinikov states \cite{IZ}.
In addition, the 2D Fermi gases with SOC can exhibit $p$-wave character in
the helicity bases. If a Zeeman field in Hamiltonian (\ref{BDG}) is added, a
novel topological phase transition from non-topological superfluids to
topological superfluids can be induced. In the topological superfluid phase,
there exist Majorana fermions and the associated non-Abelian statistics,
which are the critical ingredients for implementing topological quantum
computing \cite{Zhang}.

\subsubsection{Superfluid order gaps}

The effects of SOC are more interesting for the superfluid pairing gap $%
\Delta $. Applying the same procedure as that for the chemical potential, we
find
\begin{equation}
\Delta ^{2}\simeq 2E_{b}E_{F}+\frac{16E_{b}E_{F}}{3(E_{b}+2E_{F})^{2}}\eta
^{2}.  \label{MSTG}
\end{equation}%
There is no first-order correction with respect to $\eta $ ($\sim \alpha ^{2}
$) for $\Delta ^{2}$, and the second-order $\eta ^{2}$ ($\sim \alpha ^{4}$)
is the leading correction. Moreover, the second-order coefficient $\partial
^{2}\Delta ^{2}/\partial \eta ^{2}$ is always positive and has a maximum $%
4\eta ^{2}/3$ ($=m^{2}\alpha ^{4}/3$) when $E_{b}=2E_{F}$. The high-order
coefficients for $\eta ^{3}$ and $\eta ^{4}$ are $\Lambda
_{3}=-512E_{b}E_{F}(E_{b}-E_{F})/[15(E_{b}+2E_{F})^{4}]$ and $\Lambda
_{4}=64E_{b}E_{F}(1017E_{b}^{2}-3508E_{b}E_{F}+1076E_{F}^{2})/[315(E_{b}+2E_{F})^{6}]
$ respectively.

In order to see the effect of SOC on the superfluid pairing gap more
clearly, we introduce a dimensionless quantity
\begin{equation}
\Delta _{d}=\frac{\Delta }{\Delta _{0}}\simeq \sqrt{1+\frac{8}{%
3(E_{b}+2E_{F})^{2}}\eta ^{2}},  \label{GN}
\end{equation}%
where $\Delta _{0}$ is the superfluid pairing gap without SOC. In the
asymptotic BCS limit with a weak bound state ($E_{b}\ll E_{F}$), $\Delta
_{d}\simeq \sqrt{1+2\eta ^{2}/E_{F}^{2}{}}$. For a weak SOC $(\eta \ll
E_{F}) $ in typical solid-state materials, $\Delta _{d}\simeq 1$, which
means that the SOC does not affect the superfluid pairing gap significantly.
However, for a strong SOC $\eta \sim E_{F}$ that has been achieved for
ultracold atoms \cite{LYJ1}, this gap can be enhanced greatly. In the deep
BEC regime with a strong binding energy ($E_{b}\gg E_{F}$ and $E_{b}\gg \eta
$), $\Delta _{d}\simeq \sqrt{1+8\eta ^{2}/(3E_{b}^{2}{})}$, therefore the
superfluid pairing gap increases only slightly with the increasing SOC
strength. These analytic results agree well with the numerical results shown
in Fig. 1b. Note that similar behavior for $\Delta _{d}$ is also observed in
the numerical results in 3D \cite{3DF}.

\subsubsection{Ground-state energy per atom}

In terms of Eqs. (\ref{MSTC}) and (\ref{MSTG}), the ground-state energy per
Fermi atom $E=E_{G}/n$ can be obtained
\begin{equation}
E\simeq -\frac{1}{2}E_{F}+\frac{8E_{F}}{3(E_{b}+2E_{F})^{2}}\eta ^{2}.
\label{EG}
\end{equation}%
The comparison of Eq. (\ref{EG}) with the direct numerical simulation
results is shown in Fig. 1c. The ground-state energy, like the superfluid
pairing gap, depends on $\eta ^{2}$ ($\sim \alpha ^{4}$). In the asymptotic
BCS limit ($E_{b}\ll E_{F}$), $E\simeq -E_{F}/2+2\eta ^{2}/(3E_{F})$, which
means that the ground-state energy can be enhanced significantly only for a
large $\eta $. In the deep BEC regime ($E_{b}\gg E_{F}$ and $E_{b}\gg \eta $%
), $E\simeq -E_{F}/2+8E_{F}\eta ^{2}/(3E_{b}^{2})$, therefore the
ground-state energy only increases slightly even when $\eta \sim E_{F}$. In
addition, the high-order coefficients for $\eta ^{3}$ and $\eta ^{4}$ are
given by $E_{3}=-128E_{b}E_{F}/[5(E_{b}+2E_{F})^{4}]$ and $%
E_{4}=128E_{b}E_{F}(491E_{b}-538E_{F})/[315(E_{b}+2E_{F})^{6}]$
respectively. Note that although such mean field theory gives
qualitatively correct results, it may not agree quantitatively with
the experimental results in the deep BEC regime as shown in recent
Monte Carlo numerical simulation for the 2D Fermi gases without SOC
\cite{Bertaina}.

Finally, we want to remark that if the high-order terms in the chemical
potential $\mu $, the superfluid pairing gap $\Delta $, and the ground-state
energy per Fermi atom $E=E_{G}/n$ are included, the analytical results in
Eqs. (\ref{MSTC}), (\ref{GN}) and (\ref{EG}) fit better with the numerical
results even for $E_{F}<\alpha K_{F}<3.0E_{F}$.

\section{Discussion and Conclusion}

The mean field zero temperature BCS-BEC crossover physics discussed above
provides the first critical step towards the understanding of the 2D
spin-orbit coupled degenerate Fermi gases. Clearly, many issues need be
further explored in the future, as demonstrated by the development along
this direction after the initial submission of our paper. In particular, the
finite temperature effect need be taken into account in a realistic
experiment. Without SOC, it is known that there are no superfluids for 2D
degenerate Fermi gas at a finite temperature \cite{PCH}, where relevant
physics is the Berezinskii-Kosterlitz-Thouless (BKT) transition \cite{VLB},
leading to the generation of the vortex-antivortex pairs. Recent some
interesting effects of the SOC on the BKT transition \cite{LH} has been
investigated. In experiments, the Fermi gases are confined in a 2D harmonic
trap, whose effects may be taken into account using the local density
approximation, as shown in the recent preprint \cite{LH}. The trap geometry
for the Fermi gases may also be quais-2D instead of the strict 2D. In this
case, the confinement along the third direction may affect the critical
transition temperature, which need be further explored.

In summary, motivated by the recent experimental breakthrough for the
realization of the SOC for cold atoms and the 2D degenerate Fermi gas, we
investigate the zero temperature BCS-BEC crossover physics in 2D spin-orbit
coupled degenerate Fermi gases using the mean field approximation. By
solving the corresponding gap and atom number equations both numerically and
analytically, we reveal the ground state properties of the spin-orbit
coupled 2D Fermi gases. Our analytic results agree quantitatively with our
numerical results in 2D and qualitatively with previous numerical results
for the 3D spin-orbit coupled Fermi gases, where analytic results are
lacked. The analytic expressions for various physical quantities may provide
a powerful tool for engineering and probing many new topological phenomena
in 2D Fermi gases, including the intriguing Majorana physics and the
associated non-Abelian statistics.

\textbf{Acknowledgements}

We thank Yongping Zhang, Li Mao, and Wei Yi for helpful discussion. This
work is supported by DARPA-YFA (N66001-10-1-4025), ARO (W911NF-09-1-0248),
NSF (PHY-1104546), and AFOSR (FA9550-11-1-0313). GC is also supported by the
973 program under Grant No. 2012CB921603, the NNSFC under Grant No.
11074154, and the ZJNSF under Grant No. Y6090001.

\end{document}